\documentclass[aps,pra,preprint,showpacs]{revtex4-1}
\usepackage{amsmath}
\usepackage{amssymb}
\usepackage{graphicx}

\begin{document}

\title{Quantum spin/valley Hall effect and topological
insulator phase transitions in silicene}
\author{M. Tahir$^1$, A. Manchon$^1$, K. Sabeeh$^2$, and U. Schwingenschl\"{o}gl$^{1,}$}
\email{udo.schwingenschlogl@kaust.edu.sa,00966(0)544700080}
\affiliation{$^1$PSE Division, KAUST, Thuwal 23955-6900, Kingdom of Saudi Arabia}
\affiliation{$^2$Department of Physics, Quaid-i-Azam University, Islamabad 45320, Pakistan}

\begin{abstract}
We present a theoretical realization of quantum spin and quantum valley Hall effects
in silicene. We show that combination of an electric field and intrinsic spin-orbit
interaction leads to quantum phase transitions at the charge neutrality point. This
phase transition from a two dimensional topological insulator to a trivial
insulating state is accompanied by a quenching of the quantum spin Hall
effect and the onset of a quantum valley Hall effect, providing a tool to
experimentally tune the topological state of silicene. In contrast to
graphene and other conventional topological insulators, the proposed effects
in silicene are accessible to experiments.
\end{abstract}


\maketitle

Silicene is a single layer of silicon \cite{1,2} isostructural to graphene 
\cite{3}. It has a low buckled honeycomb structure, where the charge
carriers behave like massless Dirac fermions \cite{4}. Experimental
realizations of silicene sheets \cite{4,5} and ribbons \cite{6,7} have been
demonstrated by synthesis on metal surfaces. It is believed that silicene
can be the starting point of new opportunities for electrically tunable
electronic devices, in alternative to graphene \cite{8}. Though graphene
possesses extraordinary properties \cite{3}, its application in device
fabrication is limited by the zero band gap and the difficulty to tune Dirac
particles electrically. Moreover, even if a gap can be introduced in
graphene by chemical doping it is incompatible with existing
nanoelectronics. In the desire to overcome this limitation, the buckling in
silicene offers new possibilities for manipulating the particle dispersion
to achieve an electrically tunable band gap. In addition to the electric
field controlled gap, silicene has a relatively large intrinsic spin-orbit
interaction (SOI) induced gap of 1.55 meV \cite{9}, which provides a mass to
the Dirac fermions. This mass can be controlled experimentally by applying
an external perpendicular electric field. Recently, the SOI gap in silicene
has been studied using density functional band structure calculations \cite{9},
predicting a quantum spin Hall effect (QSHE) in an experimentally
accessible temperature regime, in contrast to graphene \cite{10,11,12,13}.

The QSHE has attracted significant interest in condensed matter physics as
it constitutes a new phase of matter \cite{10,14,15,16,17,18,19,20,21a}.
Therefore, when Kane and Mele \cite{10} in a ground breaking study of
graphene proposed a new class of insulators, the topological insulators,
a great theoretical \cite{17,18,19,20,21a} and
experimental excitement was generated \cite{15,16,19}. These materials are
insulating in the bulk, whereas the surface states are conducting and
protected against scattering by time reversal symmetry. The surface states
are chiral and have a well defined spin texture with a single Dirac cone
energy spectrum. The occurrence of conducting surface states is related to
SOI \cite{19}, which is a crucial criterion for realizing the QSHE. It was
proposed to search for new materials with strong SOI for the fabrication of
spintronic devices \cite{10,19}. In addition to the QSHE, an analogous quantum valley
Hall effect (QVHE) \cite{21a,21b} arises from a broken inversion symmetry, where
Dirac fermions in different valleys flow to opposite transverse edges when
an in-plane electric field is applied in presence of intrinsic SOI. The QVHE
paves the way to electric generation and detection of valley polarization.

In the light of the above discussion, silicene is likely to show significant
signatures of QSHE and QVHE, as well as a transition from a topologically
trivial to a band insulating state and further to a semimetallic state. The
QSHE and QVHE are the fundamental transport processes arising from intrinsic
SOI and an external perpendicular electric field, respectively. The latter
can give rise to a topologically nontrivial band insulator or semimetal,
leading to a quantized Hall and valley conductivity without magnetic field.
Here we show that the quantum phase transition from a two dimensional
topological insulator to a trivial insulating state is accompanied by a
quenching of the QSHE and the onset of a QVHE,
providing a tool to experimentally tune the topological state of
silicene. We consider silicene in the presence of an external electric field
and intrinsic SOI to discuss these phenomena, disregarding the extrinsic
SOI as it is weak (0.7 meV) with respect to the intrinsic SOI (3.9 meV) 
\cite{22,23}. We employ the standard Kubo formalism to derive the Hall
conductivity, which is experimentally accessible at reasonable temperatures
due to the remarkable buckling and strong SOI in silicene. We believe that
our results open new experimental directions for the realization of the QSHE,
QVHE, and topological insulators.

Dirac fermions in buckled silicene obey the two-dimensional graphene-like
Hamiltonian \cite{8,22,23}
\begin{equation}
H_{s_{z}}^{\eta }=v(\sigma _{x}p_{x}-\eta \sigma _{y}p_{y})-\eta s_{z}\Delta
_{SO}\sigma _{z}+\Delta _{z}\sigma _{z}.  \label{1}
\end{equation}
Here, $\eta =+/-$ for $K/K'$, $\Delta _{z}=lE_{z}$, where $E_{z}$
is the uniform electric field applied perpendicular to the silicene sheet
with $l=0.23$ \AA, ($\sigma_{x}$, $\sigma_{y}$, $\sigma_{z}$) is the
vector of Pauli matrices, and $v$ denotes the Fermi velocity of the Dirac
fermions. Spin up ($\uparrow$) and down ($\downarrow$) is represented by
$s_{z}=+1$ and $-1$, respectively. After diagonalizing the Hamiltonian given
in Eq.\ (1), we obtain the eigenvalues
\begin{equation}
E_{n,s_{z}}^{\eta }=n\sqrt{(v\hslash k)^{2}+(\Delta _{SO}-\eta s_{z}\Delta
_{z})^{2}}.  \label{2}
\end{equation}
Here, $n=+/-$ denotes the electron/hole band, and the absolute value of the
wave vector is given by $k=\sqrt{k_{x}^{2}+k_{y}^{2}}$. The corresponding
eigenfunctions for the K point with spin up are
\begin{equation}
\Psi _{+,\uparrow }^{K}=\exp [ik_{x}x+ik_{y}y]\binom{\sin \theta
/2e^{+i\varphi }}{\cos \theta /2}  \label{3}
\end{equation}
and
\begin{equation}
\Psi _{-,\uparrow }^{K}=\exp [ik_{x}x+ik_{y}y]\binom{-\cos \theta
/2e^{+i\varphi }}{\sin \theta /2}  \label{4}
\end{equation}
with $\theta =\tan^{-1}\frac{v_{F}\hslash k}{(\Delta _{SO}-\eta s_{z}\Delta
_{z})}$ and $\varphi =\tan ^{-1}\frac{k_{y}}{k_{x}}$. The corresponding
solutions for the K point with spin down are
\begin{equation}
\Psi _{+,\downarrow }^{K}=\exp [ik_{x}x+ik_{y}y]\binom{\cos \theta
/2e^{+i\varphi }}{\sin \theta /2}  \label{5}
\end{equation}
and
\begin{equation}
\Psi _{-,\downarrow }^{K}=\exp [ik_{x}x+ik_{y}y]\binom{-\sin \theta
/2e^{+i\varphi }}{\cos \theta /2}  \label{6}
\end{equation}
The eigenfunctions for the $K'$ point can be obtained by
exchanging spin up and down in the $K$ point solution with $e^{+i\varphi}$
replaced by $e^{-i\varphi}$.

First we discuss the energy eigenvalues obtained in Eq.\ (2) for the $K$ point
to explore the band splitting and quantum phase transitions. The energy is
plotted as a function of dimensionless wave number ($ka/\pi$) in Fig.\ 1 for
a lattice constant of $a=3.86$ \AA. In Fig.\ 1(a) we find a well resolved
energy gap for either the electric field or the SOI finite. This confirms a
metal to insulator transition. Figure 1(b) for finite SOI and perpendicular
external electric field with $\Delta_{SO}>\Delta_{z}$ shows an energy
splitting between the spin up and down bands of both the electrons and
holes. This splitting is less than the energy gap between the electrons and
holes. The situation reflects a topological insulating state. Figure 1(c) is
analogous to Fig.\ 1(b) but for $\Delta_{SO}<\Delta_{z}$. The splitting of
the spin up bands remains the same as in Fig.\ 1(b), while the splitting of
the spin down bands increases. This situation corresponds to a band
insulator. Figure 1(d) is analogous to Figs.\ 1(b) and (c) but for
$\Delta_{SO}=\Delta_{z}$. We see that the energy gap closes between the spin
up bands, while the spin down bands maintain a finite energy gap. This
reflects a semimetallic state \cite{24,25,26,27}. Note that for the $K'$ point the
eigenvalues are identical to those of the $K$ point if spin up and down are
exchanged.

The Hall conductivity $\sigma _{xy}$ can be obtained by the standard
Kubo formula as \cite{10,20}
\begin{equation}
\sigma _{xy}(\eta ,s_{z})=\frac{i\hslash e^{2}}{L_{x}L_{y}}\sum_{k}\frac{%
f(E_{+,s_{z}}^{\eta })-f(E_{-,s_{z}}^{\eta })}{(E_{+,s_{z}}^{\eta
}-E_{-,s_{z}}^{\eta })^{2}}[\left\langle \Psi _{-,s_{z}}^{\eta }\right\vert
v_{y}\left\vert \Psi _{+,s_{z}}^{\eta }\right\rangle \left\langle \Psi
_{+,s_{z}}^{\eta }\right\vert v_{x}\left\vert \Psi _{-,s_{z}}^{\eta
}\right\rangle ].  \label{7}
\end{equation}
The velocity components $v_{x}=v\sigma _{x}$ and $v_{y}=-\eta v\sigma _{y}$
can be obtained from the Hamiltonian in Eq.\ (1). Using Eqs.\ (2) to (6), the
expectation values of the velocities are derived and used in Eq.\ (7). We
arrive at 
\begin{equation}
\sigma _{xy}(\eta ,s_{z})=-s_{z}\frac{\hslash e^{2}v^{2}}{4(2\pi )^{2}}\int
d^{2}k\frac{(\Delta _{SO}-\eta s_{z}\Delta _{z})}{[v^2\hslash^2k^2+(\Delta
_{SO}-\eta s_{z}\Delta _{z})^{2}]^{3/2}}[f(E_{+,s_{z}}^{\eta
})-f(E_{-,s_{z}}^{\eta })].  \label{8}
\end{equation}
If the Fermi level lies in the gap we can write the conductivity given in
Eq.\ (8) for zero temperature as
\begin{equation}
\sigma _{xy}(\eta ,s_{z})=-s_{z}\frac{e^{2}}{4\hslash }\frac{\hslash
^{2}v^{2}}{(2\pi )^{2}}\underset{0}{\overset{\infty }{\int }}d^{2}k\frac{%
(\Delta _{SO}-\eta ,s_{z}\Delta _{z})}{[v^2\hslash^2k^2+(\Delta _{SO}-\eta
,s_{z}\Delta _{z})^{2}]^{3/2}}.  \label{9}
\end{equation}%
Evaluating the integral in Eq.\ (9), we obtain the spin and valley Hall
conductivities as 
\begin{eqnarray}
\sigma _{xy}(\text{Spin})&=&[\sigma _{xy}(K,\uparrow )-\sigma
_{xy}(K,\downarrow )]+[\sigma _{xy}(K^{^{\prime }},\uparrow )-\sigma
_{xy}(K^{^{\prime }},\downarrow )]  \label{10} \\
&=&-\frac{e^{2}}{2h}[\text{sgn}(\Delta _{SO}-\Delta _{z})+\text{sgn}(\Delta
_{SO}+\Delta _{z})]  \notag
\end{eqnarray}
\begin{eqnarray}
\sigma _{xy}(\text{Valley}) &=&[\sigma _{xy}(K,\uparrow )+\sigma
_{xy}(K,\downarrow )]-[\sigma _{xy}(K^{^{\prime }},\uparrow )+\sigma
_{xy}(K^{^{\prime }},\downarrow )]  \label{11} \\
&=&\frac{e^{2}}{2h}[\text{sgn}(\Delta _{SO}+\Delta _{z})-\text{sgn}(\Delta
_{SO}-\Delta _{z})].  \notag
\end{eqnarray}
From Eqs.\ (10) and (11) we see that the Hall conductivity for the $K$
and $K'$ points is not degenerate. This confirms the lifting of
the valley ($K$,$K'$) degeneracy (known for graphene) in silicene 
\cite{10,19,20}. In conventional insulators there is only a single valley.
When $\Delta _{SO}>\Delta _{z}>0$ then $\sigma
_{xy}(\text{Spin})=e^{2}/h$ and $\sigma_{xy}(\text{Valley})=0$, which
corresponds to the topological insulating state \cite{10}. When
$\Delta _{z}>\Delta _{SO}>0$ then $\sigma_{xy}(\text{Spin})=0$ and
$\sigma_{xy}(\text{Valley})=e^{2}/h$, which corresponds to a trivial
insulator \cite{21a}. This means that there is a finite Hall
conductivity even when the Fermi level is inside the band gap, implying the
presence of gapless helical edge states. This situation can be regarded as a
quantum spin Hall state. On the other hand, when the signs of $\Delta _{SO}$
and $\Delta _{z}$ are different then $\sigma _{xy}(\text{Spin})=\sigma _{xy}(\text{Valley})=0$
and there are no edge states. This corresponds to the case of a topologically trivial
insulator.

We turn to the situation when the Fermi level is in the valence or
conduction band. Now, Eq.\ (8) yields
\begin{equation}
\sigma _{xy}(\eta ,s_{z})=-s_{z}\frac{e^{2}}{4\hslash }\frac{\hslash
^{2}v^{2}}{(2\pi )^{2}}\underset{0}{\overset{k_{F}}{\int }}d^{2}k\frac{%
\Delta _{SO}-\eta s_{z}\Delta _{z}}{[v^2\hslash^2k^2+(\Delta _{SO}-\eta
s_{z}\Delta _{z})^{2}]^{3/2}}.  \label{12}
\end{equation}%
After performing the integral in Eq.\ (12) and applying the integration
limit we obtain
\begin{equation}
\sigma _{xy}(\eta ,s_{z})=-s_{z}\frac{e^{2}}{4h}[\text{sgn}(\Delta
_{SO}-\eta s_{z}\Delta _{z})-\frac{\Delta _{SO}-\eta s_{z}\Delta _{z}}{%
\sqrt{v^2\hslash^2k_{F}^{2}+(\Delta _{SO}-\eta s_{z}\Delta _{z})^{2}}}].
\label{13}
\end{equation}
Finally, the spin and valley Hall conductivities for the Fermi level in the
conduction band at zero temperature can be written as
\begin{equation}
\sigma _{xy}(\text{Spin})=-\frac{e^{2}}{2h}[\frac{\Delta _{SO}-\Delta _{z}}{\sqrt{v^2\hslash
^2k_{F}^{2}+(\Delta _{SO}-\Delta _{z})^{2}}}+\frac{\Delta _{SO}+\Delta
_{z}}{\sqrt{v^2\hslash^2k_{F}^{2}+(\Delta _{SO}+\Delta _{z})^{2}}}] 
\end{equation}
\begin{equation}
\sigma _{xy}(\text{Valley})=\frac{e^{2}}{2h}[\frac{\Delta _{SO}+\Delta _{z}}{\sqrt{v^2\hslash
^2k_{F}^{2}+(\Delta _{SO}+\Delta _{z})^{2}}}-\frac{\Delta _{SO}-\Delta
_{z}}{\sqrt{v^2\hslash^2k_{F}^{2}+(\Delta _{SO}-\Delta _{z})^{2}}}] 
\end{equation}
Equations (14) and (15) show that the $K$ and $K'$ points are not degenerate.  
We obtain similar results for the case when the Fermi level lies in the
valence band due to the electron-hole symmetry.
A closer analytical examination of Eqs.\ (14) and (15) implies that silicene
is subject to quantum phase transitions as shown and discussed in Fig.\ 1.

We show in Fig.\ 2 the quantum spin and valley Hall conductivities
as functions of the electric field for a fixed value of the SOI.
By tuning the electric field one may tune the SHE and VHE. The transition
is striking in the case of the QSHE and QVHE (i.e., the Fermi level is in
the gap). However, also in the realistic case when the Fermi level is
above or below the gap it is possible to observe this topological
transition. For comparison with existing work, if we assume that the
perpendicular electric field is zero ($\Delta _{z}=0$) our results are
similar to those obtained for the QSHE in graphene and conventional systems 
\cite{10,19,20}. However, in silicene the SOI is much stronger and should be
accessible experimentally at reasonable temperatures. Further, with the
inclusion of $\Delta _{z}$ the SOI becomes more pronounced, giving rise to a
finite mass term. The opposite signs of the $\Delta _{z}\sigma _{z}$ terms
in the spin up and down channels of the effective Hamiltonian lift the spin
and valley degeneracies. This distinguishes silicene from
graphene and is responsible for the quantum phase transitions from a
topological insulator to a band insulator and further to a
semimetal. Such a general mechanism is not possible in graphene and
conventional topological insulators with perpendicular electric field, while
it is guaranteed in silicene by the two opposite spins and valleys.

The Hamiltonian in Eq.\ (1) can also be used to describe germanene, which is
a honeycomb structure of germanium \cite{4,9,22}. Here the SOI is even
stronger ($\Delta_{SO}=43$ meV) with $l=0.33$ \AA. Hence, the above
analysis is fully applicable to germanene. Our results imply that the SOI
splitting can be controlled by a gate voltage or external electric field.
This is of significance for electrically tunable spintronics devices
\cite{20}, especially Datta-Das spin field-effect transistors \cite{28}.
Therefore, realization of the QSHE, QVHE, and quantum phase transitions in
silicene and germanene topological insulators provides new spintronic
materials for high efficiency electric spin manipulation.

To conclude, we have carried out analytical calculations for the Hall
conductivity in silicene, based on the Kubo formula. An electric field is
included to take into account the effect of electrical tuning for
nanoelectronic applications. In an insulator where the chemical potential is
located inside the energy gap between the conduction and valence bands the
Hall conductivity is an integer multiple of $e^{2}/h$. The calculated Hall
conductivity of silicene jumps from 1 to 0 at the critical electric field of
the level crossing. Experimentally, the best way to see the QSHE and QVHE is
to measure the Hall conductivity as a function of the gate voltage (that
tunes the chemical potential). A plateau in the Hall conductivity should be
observed when the chemical potential is inside the gap. In silicene and
germanene, the temperature will not affect the Hall plateau as the SOI is
strong. It is shown that the energy splitting due to the SOI and
electric field leads to QSHE and QVHE with quantum phase transitions at the
charge neutrality point. Thus QSHE, QVHE, and the topological insulating states
in silicene and germanene can be observed experimentally at finite temperature.

\begin{figure*}[b]
\includegraphics[width=1\textwidth,clip]{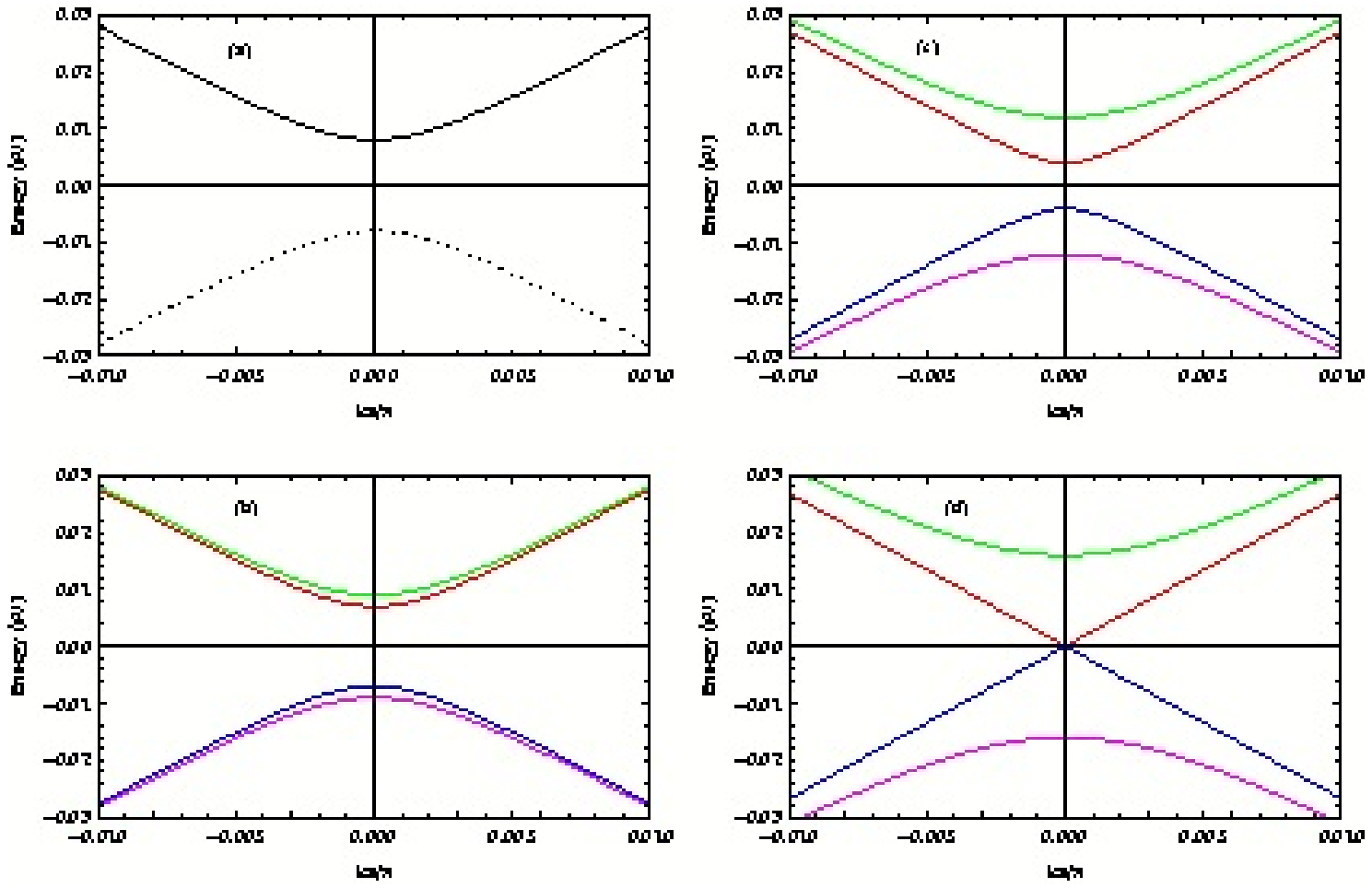}
\caption{Energy as a function of dimensionless wave number for fixed
SOI ($\Delta_{so}$), perpendicular electric field ($\Delta_{z}$), and
velocity $v=5\times10^{5}$ m/s. Red: spin up electron, Green: spin down
electron, Blue: spin up hole, and Magenta: spin down hole.
(a) Either $\Delta_{so}=0$ meV and $\Delta_{z}=8$ meV
or $\Delta_{so}=8$ meV and $\Delta_{z}=0$ meV. (b) $\Delta_{so}=8$
meV and $\Delta_{z}=4$ meV. (c) $\Delta_{so}=4$ meV and $\Delta_{z}=8$ meV.
(d) $\Delta_{so}=\Delta_{z}=8$ meV.} 
\end{figure*}

\begin{figure}
\includegraphics[width=0.5\textwidth,clip]{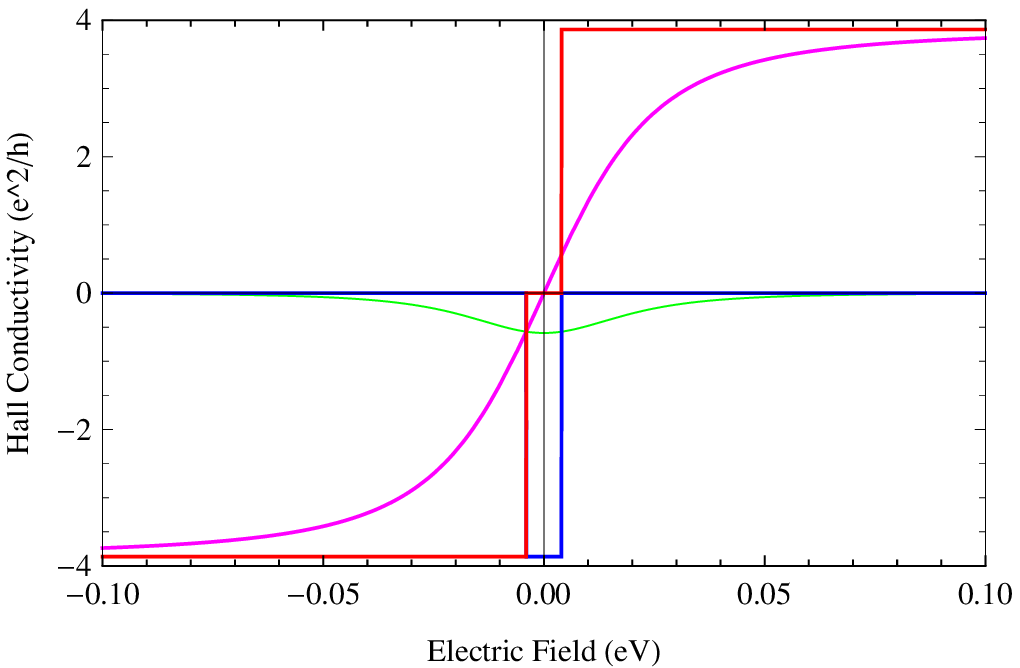}
\caption{Quantum spin and valley Hall conductivities as a function of the
perpendicular electric field $\Delta _{z}$ for a fixed value of the SOI
(3.9 meV). Blue and green: Quantum spin Hall conductivities for the
Fermi level in the band gap and the conduction band, respectively. Red and
magenta: Valley Hall conductivities for the Fermi level
in the band gap and the conduction band, respectively.} 
\end{figure}


\begin{thebibliography}{99}
\bibitem{1} K. Takeda and K. Shiraishi, Phys. Rev. B 50, 14916 (1994).

\bibitem{2} G.\thinspace G. Guzm\'{a}n-Verri and L.\thinspace C. Lew Yan
Voon, Phys. Rev. B 76, 075131 (2007).

\bibitem{3} A.\thinspace K. Geim, Science 324, 1530 (2009); A. H. Castro
Neto, F. Guinea, N. M. R. Peres, K. S. Novoselov, and A. K. Geim, Rev. Mod.
Phys. 81, 109 (2009).

\bibitem{4} P. Vogt, P. D. Padova, C. Quaresima, J. Avila, E. Frantzeskakis,
M. C. Asensio, A. Resta, B. Ealet, and G. L. Lay, Phys. Rev. Lett. 108,
155501 (2012).

\bibitem{5} B. Lalmi, H. Oughaddou, H. Enriquez, A. Kara, S. Vizzini, B.
Ealet, and B. Aufray, Appl. Phys. Lett. 97, 223109 (2010).

\bibitem{6} B. Aufray, A. Kara, S. Vizzini, H. Oughaddou, C. L\'{e}andri, B.
Ealet, and G. L. Lay, Appl. Phys. Lett. 96, 183102 (2010).

\bibitem{7} P. E. Padova, C. Quaresima, C. Ottaviani, P. M. Sheverdyaeva, P.
Moras, C. Carbone, D. Topwal, B. Olivieri, A. Kara, H. Oughaddou, B. Aufray,
and G. L. Lay, Appl. Phys. Lett. 96, 261905 (2010).

\bibitem{8} N. D. Drummond, V. Z\'{o}lyomi, and V. I. Fal'ko, Phys. Rev. B
85, 075423 (2012).

\bibitem{9} C. C. Liu, W. Feng, and Y. Yao, Phys. Rev. Lett. 107, 076802
(2011).

\bibitem{10} C. L. Kane and E. J. Mele, Phys. Rev. Lett.
95, 226801 (2005).

\bibitem{11} Y.\thinspace G. Yao, F. Ye, X.\thinspace L. Qi, S.\thinspace C.
Zhang, and Z. Fang, Phys. Rev. B 75, 041401(R) (2007).

\bibitem{12} D. Huertas-Hernando, F. Guinea, and A. Brataas, Phys. Rev. B
74, 155426 (2006).

\bibitem{13} H. Min, J.\thinspace E. Hill, N.\thinspace A. Sinitsyn,
B.\thinspace R. Sahu, and L. Kleinman, A.\thinspace H. MacDonald, Phys. Rev.
B 74, 165310 (2006).

\bibitem{14} C.\thinspace L. Kane and E.\thinspace J. Mele, Phys. Rev. Lett.
95, 146802 (2005).

\bibitem{15} B.\thinspace A. Bernevig, T.\thinspace L. Hughes, and
S.\thinspace C. Zhang, Science 314, 1757 (2006).

\bibitem{16} M. K\"{o}nig, S. Wiedmann, C. Br\"{u}ne, A. Roth, H. Buhmann,
L.\thinspace W. Molenkamp, X.\thinspace L. Qi, and S.\thinspace C. Zhang,
Science 318, 766 (2007).

\bibitem{17} S. Murakami, Phys. Rev. Lett. 97, 236805 (2006).

\bibitem{18} C. X. Liu, T. L. Hughes, X.\thinspace L.
Qi, K. Wang, and S. C. Zhang, Phys. Rev. Lett. 100, 236601 (2008).

\bibitem{19} D. Pesin and A. H. MacDonald, Nat. Mat. 11, 409 (2012).

\bibitem{20} N. A. Sinitsyn, J. E. Hill, H. Min, J. Sinova, and A. H.
MacDonald, Phys. Rev. Lett. 97, 106804 (2006).

\bibitem{21a} D. Xiao, W. Yao, and Q. Niu, Phys. Rev. Lett. 99, 236809
(2007).

\bibitem{21b} A. Rycerz, J. Tworzydlo, and C. W. J. Beenakker, Nat. Phys. 3, 172
(2007).

\bibitem{22} C. C. Liu, H. Jiang, and Y. Yao, Phys. Rev. B 84, 195430 (2011).

\bibitem{23} M. Ezawa, New J. Phys. 14, 03303 (2012).

\bibitem{24} M. Z. Hasan and C. L. Kane, Rev. Mod. Phys. 82, 3045 (2010).

\bibitem{25} X.-L. Qi and S.-C. Zhang, Rev. Mod. Phys. 83, 1057 (2011).

\bibitem{26} Y. Xia, D. Qian, D. Hsieh, L. Wray, A. Pal, H. Lin, A. Bansil,
D. Grauer, Y. S. Hor, R. J. Cava, and M. Z. Hasan, Nat. Phys. 5, 398 (2009).

\bibitem{27} Z. Wang, Y. Sun, X. Q. Chen, C. Franchini, G. Xu, H. Weng, X.
Dai, and Z. Fang, Phys. Rev. B 85, 195320 (2012).

\bibitem{28} S. Datta and B. Das, Appl. Phys. Lett. 56, 665 (1990).
\end{thebibliography}
\end{document}